\documentclass[a4paper,11pt]{article}
\pdfoutput=1 % if your are submitting a pdflatex (i.e. if you have
\usepackage{jinstpub} % for details on the use of the package, please
\title{\boldmath Prototyping and qualification of 2S modules for the CMS Outer Tracker upgrade at the HL-LHC}

\author[a,1]{F. Zhang,\note{Corresponding author.}}
\author[b]{G. Blanchot,}
\author[c]{S. Higginbotham,}
\author[d]{A. Honma,}
\author[c]{A. Kalogeropoulos,}
\author[b]{M. Kovacs,}
\author[b]{A. La Rosa,}
\author[e]{S. Nasr}
\author{on behalf of the Tracker Group of the CMS collaboration}
\affiliation[a]{University of California, Davis, 1 Shields Avenue, U.S.A}
\affiliation[b]{CERN, Geneva 1211, Switzerland}
\affiliation[c]{Princeton University, Princeton, NJ 08544, U.S.A}
\affiliation[d]{Brown University, Providence, RI 02912, U.S.A}
\affiliation[e]{Bristol University, Bristol BS8 1TH, UK}

\emailAdd{fengwang@cern.ch}

\abstract{In preparation for the High Luminosity LHC, the entire tracker detector of the CMS experiment will be exchanged as part of the Phase-2 Upgrade. The new Outer Tracker will comprise approximately 13,000 silicon sensor modules, of which 7608 are ``2S modules'' consisting of two parallel mounted silicon strip sensors, and 5592 are ``PS modules'' consisting of one pixel and one strip sensor in a single module. These modules provide tracking information to the Level 1 trigger by correlating the hit information of both sensor layers, allowing discrimination of particle tracks by their transverse momentum. To guarantee successful operation during data-taking, the production of the outer tracker modules has to fulfill strict requirements. This note will discuss the assembly procedures as well as some key results of the electronic, thermal and vibration tests performed at CERN for qualifying the 2S module design and for preparing the module assembly procedures.}

\keywords{Particle tracking detectors (Solid-state detectors)}

\proceeding{12$^{\text{th}}$ International Conference on Position Sensitive Detectors\\
  12-17 June 2021\\
  Birmingham, UK}

\begin{document}
\maketitle
\flushbottom

\section{Introduction}
\label{sec:intro}
\begin{figure}[htbp]
\centering
\includegraphics[width=0.8\textwidth]{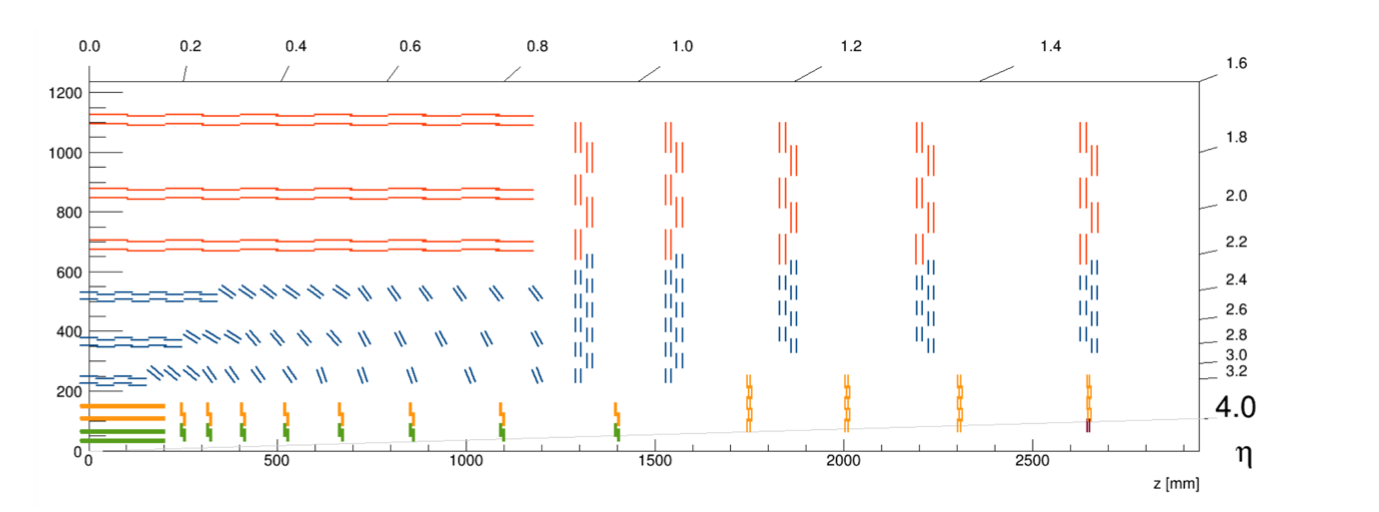}
\caption{Schematic of one quarter of the tracker layout 
in r-z dimension~\cite{e}. In the Inner Tracker, the green lines represent pixel modules 
made of two readout chips, and the orange lines represent pixel modules with four 
readout chips. In the Outer Tracker, the blue lines represent macro pixel-strip 
(PS) modules, and the red lines represent strip-strip (2S) modules.}
\label{fig:phase2tracker}
\end{figure}

The Large Hadron Collider (LHC) at CERN will be upgraded to the High 
Luminosity LHC (HL-LHC)~\cite{a} in the next few years. The instantaneous 
luminosity of the beam collisions will reach between 
5 $\times$ 10$^{34}$ cm$^{-2}$s$^{-1}$ and 7 $\times$ 10$^{34}$ cm$^{-2}$s$^{-1}$, and the integrated luminosity is expected to reach 4000 fb$^{-1}$. As one of the 
general-purpose detector in the LHC, the Compact Muon Solenoid (CMS) 
experiment~\cite{b} will need to improve its 
tracking resolution and the ability to reconstruct and select the events of 
physics interests in such luminosity conditions. One of the crucial parts 
is the silicon tracker sub-detector~\cite{b,c} (pixel + strip systems) which will need to be replaced. 
Hence, a new tracker 
detector~\cite{d} has been designed and is in the process of qualification before its full assembly. 
Comprising a total silicon detector area of 220 m$^{2}$, the system will consist of two sub-detectors: the 
Inner Tracker (IT) and Outer Tracker (OT). The IT will be composed of silicon pixel 
modules, while the OT will be made of silicon strip-strip (2S) and macro pixel-strip 
(PS) modules. Figure~\ref{fig:phase2tracker} shows the design of the new tracker detector.

\section{2S module schematic and prototype}
\begin{figure}[htbp]
\centering
\includegraphics[width=0.51\textwidth]{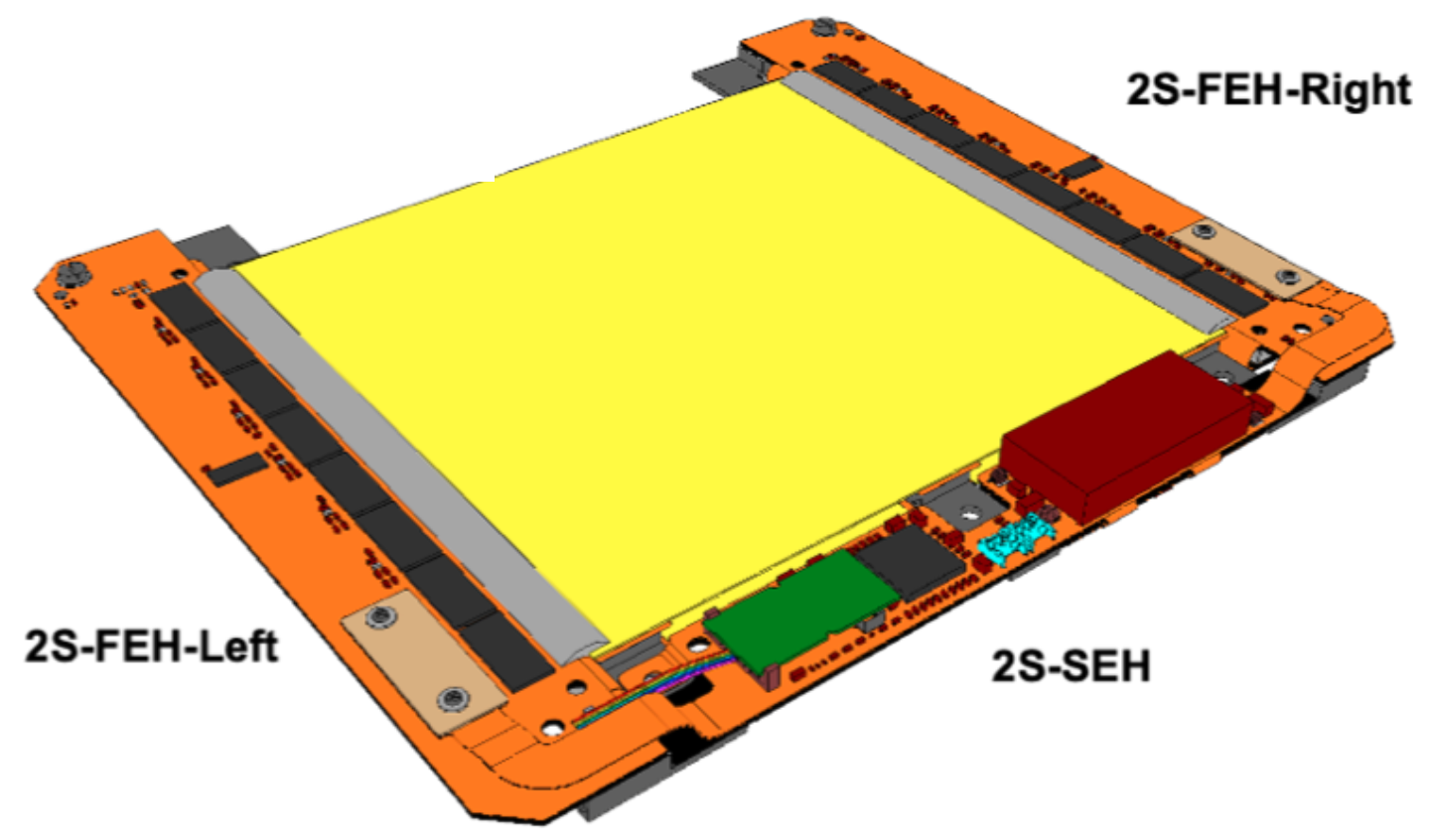}
\includegraphics[width=0.44\textwidth]{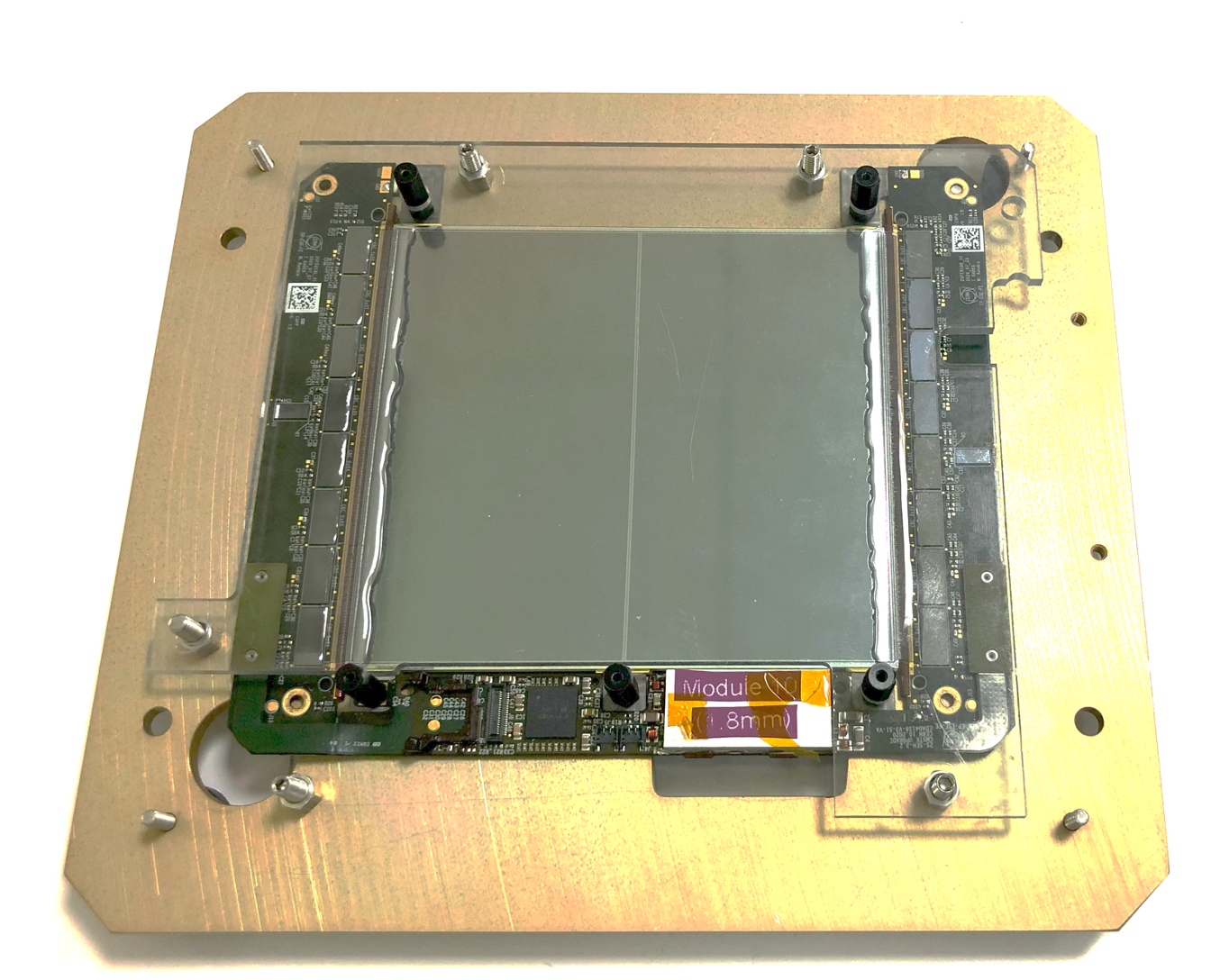}
\caption{Left: A schematic view of a 2S module. Right: A 2S module prototype.}
\label{fig:2smoduleschematic}
\end{figure}

Each 2S module is made of two silicon strip sensors placed back to back, separated with 
three aluminium-carbon fiber (AlCF) bridges and three electronic hybrids. 
Figure~\ref{fig:2smoduleschematic} shows a sketch and prototype of a 
2S module, which should be nearly identical, both mechanically and electrically, 
to the final version of 2S modules.
Each silicon sensor contains 2032 strips, each of which has a 90 $\mu$m pitch.
The gap between the two sensors has a width of 1.8 mm. Two Front-End Hybrids (FEHs) are wire-bonded with the sensors to read out the 
electric signals from them. Each of the hybrid is mounted with 8 CMS binary Chips (CBC)~\cite{d} 
and a Concentrator Integrated Circuit (CIC) chip~\cite{d}. A CBC chip reads 254 strips (127 from both the top and the bottom), performs hit correlations between the two sensors, and sends the stub data~\cite{e} out at each bunch crossing to the CIC. The CIC performs data sparsification, formats the output data, and sends them to the Service Hybrid (SEH).
The SEH is connected with the FEHs and the sensors 
to power them. It hosts a LpGBT~\cite{f}, VTRx+ optical link~\cite{g}, DC-DC converters~\cite{h} and High Voltage (HV) distribution circuitry. The data from the FEHs are merged and sent via a single optical fibre to the back-end electronic system.
The aluminium bonding wires are encapsulated 
with Sylgard 186 glue in order to protect the wires and avoid electric sparks at the 
edges between the sensors and FEHs.

The prototype assembly procedure is summarized in~\cite{e}, and consists of: gluing Kapton isolators and HV
circuitry on the backsides of the sensors; wire-bonding and encapsulating the HV circuitry; 
gluing AlCF bridges to the backsides of the sensors and make a sandwich; gluing 
hybrids to the sandwich; wire-bonding the FEHs to the sandwich and encapsulating the
wires. A total of 10 2S modules have been built at CERN and much has been learned from these that has contributed to the improvement in both the module design and the assembly tooling and procedures.

\section{Electrical and functional tests}
\begin{figure}[htbp]
\centering
\includegraphics[width=0.5\textwidth]{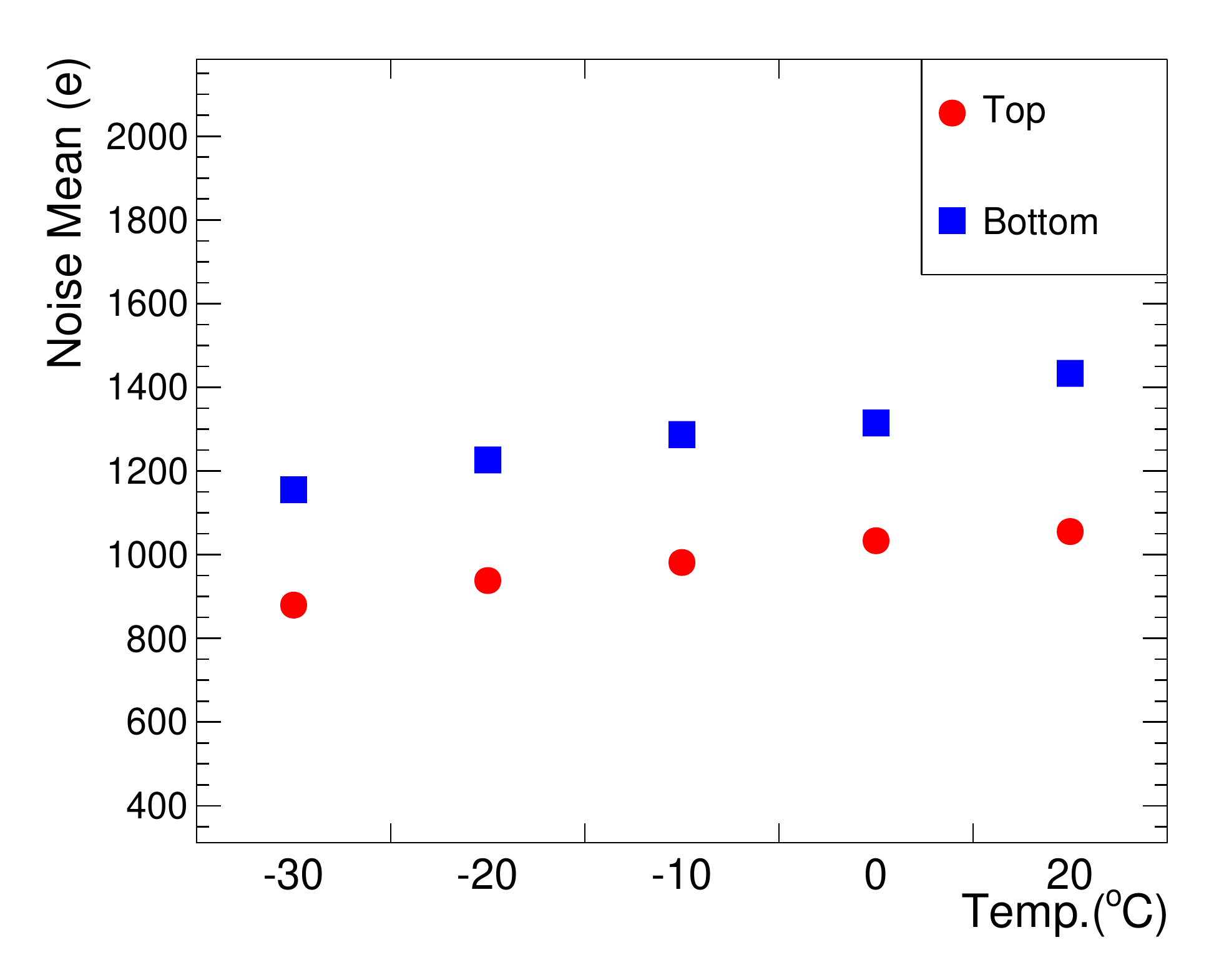}
\caption{Mean value of the noise of each side of a 2S module at different temperature 
conditions. The unit of the noise is the number of electrons.}
\label{fig:modulenoise}
\end{figure}

Noise is one of the most important features of the 2S modules. Ideally it should 
be below 1000 electrons~\cite{d} after a new 2S module is assembled, so that it will not 
swamp the signal when a particle is traversing the 
sensors from the collision center. The noise of a module will increase from radiation damage to the sensors over the lifetime of the detector. 
Therefore, a reasonable noise level is required during the prototyping and 
qualification. The module noise is also influenced by the surrounding 
temperature.
Because the 2S modules are expected to be operated at about -30$^{\circ}$C during 
the HL-LHC phase, it is important to check how mechanics and electronics will behave over the large operating temperature range between room temperature (20$^{\circ}$C) and 
-30$^{\circ}$C. 
Figure~\ref{fig:modulenoise} shows the measured mean value of noise of 
a 2S module prototype at different temperature conditions after 3 thermal 
cycles, 
separated by the top and bottom sides. The bottom side has higher noise than the top 
side because the length of the lines on the hybrid 
connecting the bottom sensor with the CBC3 chips are much longer, plus additional pick-up effects and input capacitance. The average noise of each side decreases when the temperature gets lower. The bottom side of this prototype shows a bit higher noise than the expectation, which is still being investigated.

\section{Vibration tests of module robustness}
\begin{figure}[htbp]
\centering
\includegraphics[width=0.43\textwidth]{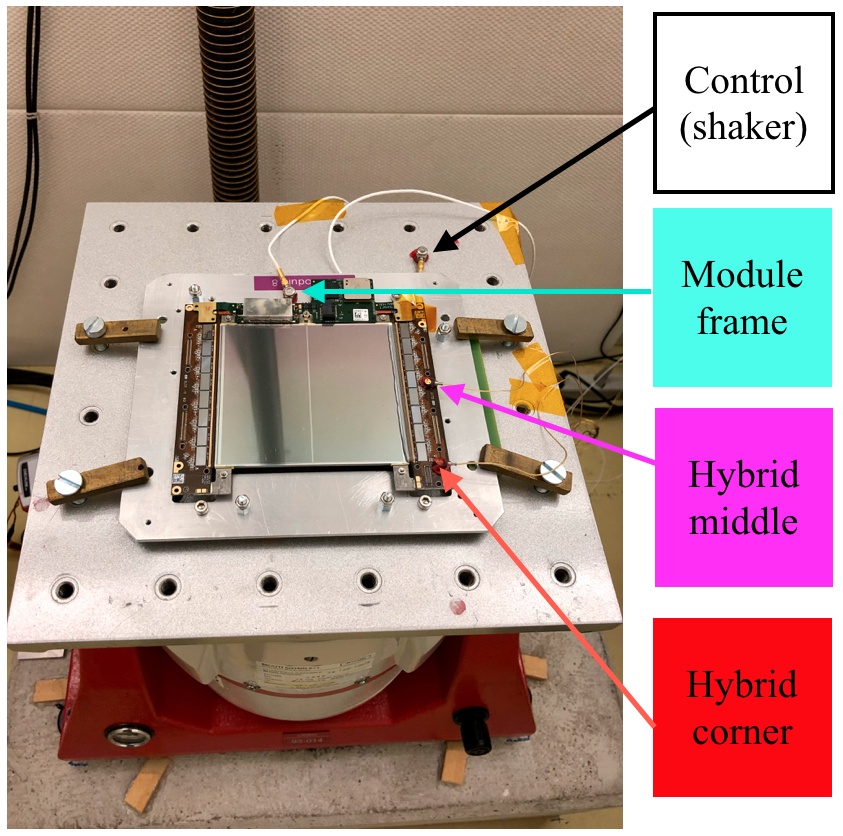}
\includegraphics[width=0.51\textwidth]{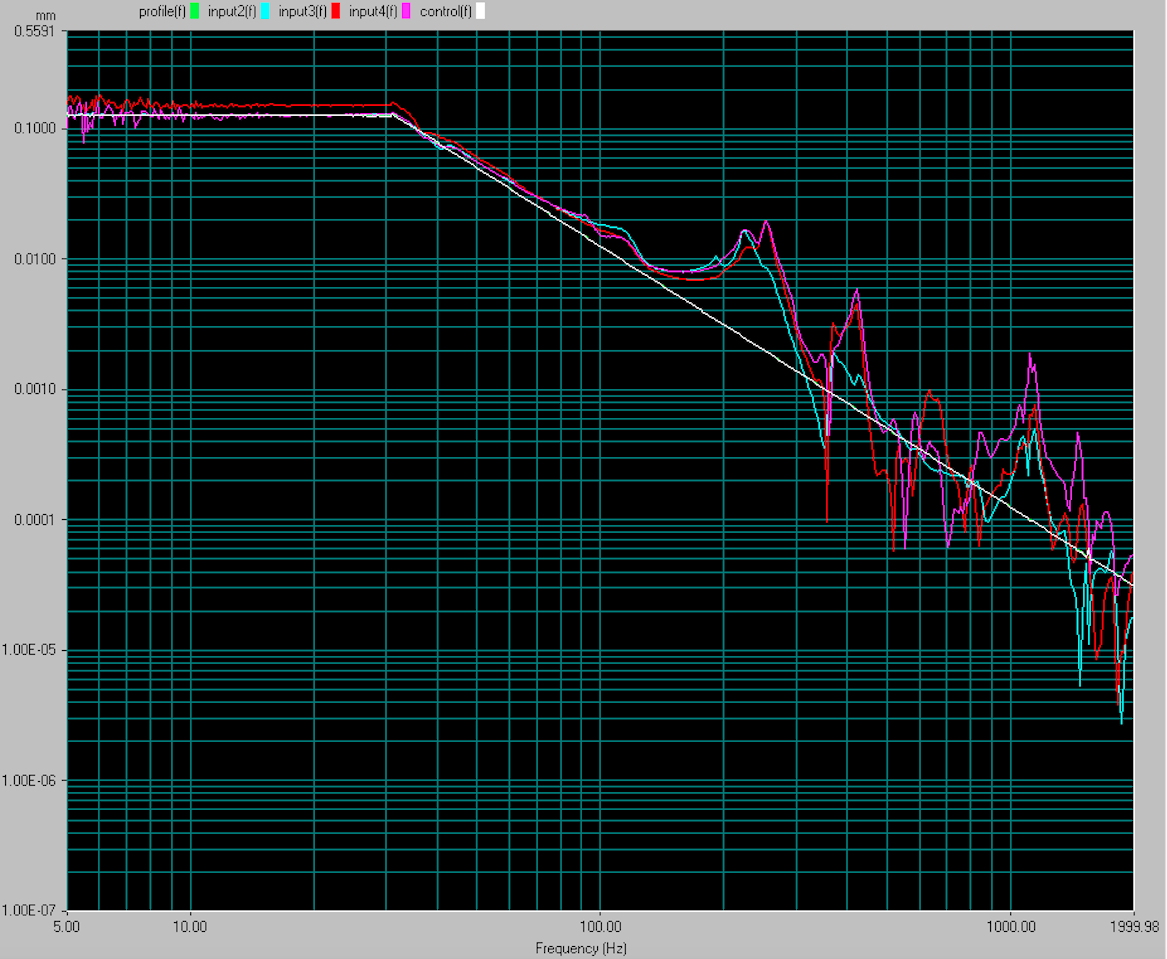}
\caption{Left: A 2S module being tested with vibration machine conducting 
resonant frequency scan. There are 4 probes placed in different positions of the 
module to measure the displacement and acceleration. Right: Displacement of 
different spots of the module as a function of the vibration frequency. Peaks 
indicate resonant frequencies of the module.}
\label{fig:vibrationtests}
\end{figure}

Since transport vibrations can cause module damage, a resonant frequency test has been conducted at the CERN EP-DT QART lab~\cite{I}. The test focuses on the movement of the hybrids with respect to the transport plate where the module is secured. 
Figure~\ref{fig:vibrationtests} shows the setup and results of a vibration test for a 2S module prototype in the vertical direction.
The module is fixed on a module carrier and placed on a high power shaker, which vibrates the module orthogonal to the plane of sensors with a sine wave frequency sweep in the range of 5-1000 Hz. 
Four probes are attached to different positions of the module. 
According to the plot, the encapsulated module has a resonance frequency 
at around 250 Hz. From the test we have confirmed the importance of encapsulating the wire bonds for their protection and adding more securing points to the hybrids to avoid damage.

\section{Summary and outlook}
At CERN a particular effort is made for developing the module assembly techniques and 
qualification process for the 2S module. %During the last years, 10 modules were built (7x 8CBC2 modules, 1x 8CBC3 module with CBC3 FEHs and CIC mezzanine, and SEH with GBTx and VTRx, and 2x final module prototypes with FEHs and SEH with lpGBT \& VTRx+) to test and improve 
%the module design and assembly procedures. 
In this note, a prototype with the latest version of hybrids has been shown, together with its schematic and the assembly procedure. The prototype has been subject to many electronic, 
electrical, thermal cycling and vibration tests in order to check that it meets the mechanical and electronic requirements and have the necessary 
robustness to survive the long working life under harsh environmental conditions expected in the HL-LHC.
The noise of the module is slightly higher than the expectation, and is reduced when the temperature decreases.
The prototype has a resonance frequency at around 250 Hz along the vertical axis, which should be damped during the transport. This module provides an important reference to the improvements made for the modules to be assembled in future, which should allow for a much smoother 
pre-production and production of the module assembly.

\acknowledgments
We would like to acknowledge the EP-DT QART lab at CERN for providing us with the vibration equipment to test one of our 
module prototypes.
%\paragraph{Note added.} This is also a good position for notes added
%after the paper has been written.

\end{document}